\documentclass[a4paper,11pt]{article}
\pdfoutput=1
\usepackage{jheppub}

\usepackage{amssymb,amsmath,amsfonts}
\usepackage[normalem]{ulem}
\usepackage[utf8x]{inputenc}
\usepackage{slashed}
\usepackage{graphicx}
\usepackage{tabularx}
\usepackage{here}
\usepackage{color}
\usepackage{csquotes} %fix backwards quotes
\usepackage{comment}
\usepackage{mathrsfs}
\usepackage{float}
\usepackage{ascmac}
\usepackage{multirow}
\usepackage{longtable}
\usepackage{bm}
\usepackage{ulem}
\usepackage{tikz}
\usepackage{tikz-feynman}
\tikzfeynmanset{compat=1.1.0, warn luatex=false}
\usetikzlibrary{decorations.pathmorphing,positioning,shapes.geometric,calc,arrows.meta}

\usepackage[italicdiff]{physics}
\newcommand{\nn}{\nonumber}

\newcommand{\figref}[1]{Fig.~\ref{#1}}

\makeatletter
\newcommand*\rel@kern[1]{\kern#1\dimexpr\macc@kerna}
\newcommand*\widebar[1]{%
  \begingroup
  \def\mathaccent##1##2{%
    \rel@kern{0.8}%
    \overline{\rel@kern{-0.8}\macc@nucleus\rel@kern{0.2}}%
    \rel@kern{-0.2}%
  }%
  \macc@depth\@ne
  \let\math@bgroup\@empty \let\math@egroup\macc@set@skewchar
  \mathsurround\z@ \frozen@everymath{\mathgroup\macc@group\relax}%
  \macc@set@skewchar\relax
  \let\mathaccentV\macc@nested@a
  \macc@nested@a\relax111{#1}%
  \endgroup
}
\makeatother

\numberwithin{equation}{section}

\preprint{
\begin{minipage}{5cm}
\small
\flushright
EPHOU-25-008\\
KYUSHU-HET-324
\end{minipage}}

\title{Radiative neutrino mass models from non-invertible selection rules}

\author{Tatsuo Kobayashi$^{1}$,} 
\author{Hiroshi Okada$^{2}$, and}
\author{Hajime Otsuka$^{3,4}$} 
\affiliation{
$^1$Department of Physics, Hokkaido University, Sapporo 060-0810, Japan}
\affiliation{
$^2$Department of Physics, Henan Normal University, Xinxiang 453007, China}
\affiliation{
$^3$Department of Physics, Kyushu University, 744 Motooka, Nishi-ku, Fukuoka 819-0395, Japan}
\affiliation{
$^4$Quantum and Spacetime Research Institute (QuaSR), Kyushu University, \\ 744 Motooka, Nishi-ku, Fukuoka 819-0395, Japan}
\emailAdd{kobayashi@particle.sci.hokudai.ac.jp}
\emailAdd{hiroshi3okada@htu.edu.cn}
\emailAdd{otsuka.hajime@phys.kyushu-u.ac.jp}

\abstract{
We apply non-invertible selection rules coming from a fusion algebra to radiative neutrino mass models where fields are labeled by the elements in the algebra. 
Since non-invertible selection rules only hold at tree level, radiative corrections naturally explain the origin of tiny neutrino masses. 
Furthermore, a remnant symmetry of the fusion algebra protects the stability of dark matter, which is conventionally imposed in radiative neutrino models. We also find that interesting neutrino mass textures are realized by assigning fields to family-dependent elements in the algebra. 
}

\makeatletter
\gdef\@fpheader{}
\makeatother

\begin{document}

\maketitle

%%%%%%%%%%%%%%%%%%%%%%%%%%%%%%%%%%%%%%%%%%%%%%%%%%%%%%%%%%%%%%%%%
\section{Introduction}

Selection rule is a fundamental principle in quantum system and quantum field theory, which determines whether couplings are allowed or forbidden. 
Conventionally, selection rules are associated with conserved laws originating from a certain group-like symmetry. 
However, it was known that non-invertible selection rules appear in quantum field theory, e.g., the tensor product of fields which are assigned as conjugacy classes or irreducible representations of a certain group. 
In addition, it also appears in the worldsheet theory of perturbative string theory \cite{Bhardwaj:2017xup}, heterotic string theory on toroidal orbifolds \cite{Dijkgraaf:1987vp,Kobayashi:2004ya, Kobayashi:2006wq,Beye:2014nxa,Thorngren:2021yso,Heckman:2024obe,Kaidi:2024wio}\footnote{The operator algebra in the  worldsheet theory on $S^1/\mathbb{Z}_2$ leads to the representation algebra of $D_4$ \cite{Dijkgraaf:1987vp,Beye:2014nxa,Thorngren:2021yso,Kaidi:2024wio,Heckman:2024obe}. Similarly, the theory on $T^2/\mathbb{Z}_3$ leads to $\Delta(54)$ \cite{Kobayashi:2006wq,Beye:2014nxa}.} and Calabi-Yau threefolds \cite{Dong:2025pah}, and type II intersecting/magnetized D-brane models \cite{Kobayashi:2024yqq,Funakoshi:2024uvy}. 

Such non-invertible selection rules are recently applied in particle physics \cite{Choi:2022jqy,Cordova:2022fhg,Cordova:2022ieu,Cordova:2024ypu,Kobayashi:2024cvp,Kobayashi:2024yqq,Kobayashi:2025znw,Suzuki:2025oov,Liang:2025dkm,Kobayashi:2025ldi}. In particular, they lead to zero textures in the Yukawa matrix, including nearest neighbor interaction texture \cite{Branco:1988iq} and its extension is found in Refs. \cite{Kobayashi:2024cvp,Kobayashi:2025znw}.
The realistic masses and mixings as well as CP phases are derived in the quark \cite{Kobayashi:2024cvp,Kobayashi:2025znw} and
lepton sectors \cite{Kobayashi:2025ldi}. 
Furthermore, it can address the strong CP problem without axion by assuming the CP invariance in ultra-violet (UV) theory \cite{Kobayashi:2025znw,Liang:2025dkm}. 
However, it was pointed out in Refs. \cite{Heckman:2024obe,Kaidi:2024wio} that the selection rules are violated at quantum levels even though they hold at tree level. 
Then, interaction terms violating tree-level selection rules are radiatively generated in the Lagrangian. 
Since they are suppressed by loop factors, there are several possibilities to explain the tiny physical observables in the Standard Model (SM) of particle physics.

In this paper, we utilize this feature to explain the origin of tiny neutrino masses.
Conventionally, we impose continuous or discrete invertible symmetries to forbid certain couplings for fields and to ensure the stability of Dark Matter (DM). 
However, such a group-like symmetry can be replaced by non-invertible selection rules, and one would test phenomenological consequences of non-invertible selection rules by high-energy experiments. 
In particular, we utilize three phenomenological features of non-invertible selection rules in the context of radiative neutrino mass models:
%%%
\begin{itemize}
\item
Even if bare interaction terms are forbidden by non-invertible selection rules, certain couplings can be generated by quantum corrections.
%%%
\item
There exists a remnant Abelian symmetry at all-loop order, so-called groupification of the fusion algebras \cite{Kaidi:2024wio}.
%%%
\item
Texture zeros of the mass matrix can be derived from the non-invertible selection rules \cite{Kobayashi:2024cvp,Kobayashi:2025znw}. 
\end{itemize}
We find that they work well for prototypical radiative neutrino mass models such as Ma model \cite{Ma:2006km}, type-II neutrino model \cite{Kanemura:2012rj}, and an extended Zee-Babu type neutrino model \cite{Zee:1985id,Babu:1988ki}.

This paper is organized as follows. 
In section \ref{sec:selection-rule}, we briefly review on non-invertible fusion algebras and their implications in field theory. 
In section \ref{sec:features}, we briefly summarize three phenomenological features of non-invertible selection rules in the context of radiative neutrino mass models. 
In section \ref{sec:examples}, we explicitly present radiative neutrino mass models by utilizing non-invertible selection rules. 
By assigning fields to the elements of fusion algebras, one can realize the texture structure for the lepton mass matrices, as discussed in section \ref{sec:Matextures}. 
Finally, section \ref{sec:con} is devoted to the conclusions. 
We summarize the non-invertible selection rules used in this paper in Appendix \ref{app}.

\section{Non-invertible selection rules}
\label{sec:selection-rule}

Here, we give a brief review on non-invertible fusion algebras.
Let us consider the fusion algebra realized by conjugacy classes or the irreducible representations of a finite group $G$. 
When we represent the elements of fusion algebra $\{a_0, a_1,a_2,...\}$ with $a_0=\mathbb{I}$ being the unit, they obey the following multiplication law:
\begin{align}
\label{eq:rules}
    a_i a_j = \sum_{k}c_{ij}^k a_k\,,
\end{align}
where the structure constants $c_{ij}^k$ are assumed to be non-negative integers.\footnote{For more general cases $c_{ij}^k$, it is called a hypergroup as discussed in Ref.~\cite{Kaidi:2024wio}.} 
As discussed in Ref.~\cite{Kaidi:2024wio}, we assign elements of fusion algebra to fields in quantum field theory, e.g., $a_i$ for a field $\phi_i$.
The tree-level couplings $\phi_1\phi_2\cdots \phi_n$ are allowed when the product of corresponding elements $a_1a_2\cdots a_n$ includes the unit $\mathbb{I}$. 
Hence, the fusion algebra leads to non-trivial selection rules for fields since the elements of the fusion algebra do not obey the multiplication law of group-like symmetries in general. However, the non-invertible selection rules hold only at the tree level.

One of the famous non-invertible algebras is the Fibonacci fusion algebra of two elements, $\mathbb{I}$ and $\tau$.
They satisfy the following algebraic relations:
\begin{align}
\label{eq:Fibo}
    \mathbb{I}\otimes \mathbb{I} = \mathbb{I},\quad
    \mathbb{I}\otimes \tau = \tau \otimes \mathbb{I} = \tau,\quad    
    \tau \otimes \tau = \mathbb{I} \oplus \tau.
\end{align}
Another one is the Ising algebra of three elements: $\mathbb{I}$ $\epsilon$, and $\sigma$.
They satisfy the following algebraic relations:
\begin{align}
\label{eq:Ising}
    \mathbb{I}\otimes \mathbb{I} = \mathbb{I},\quad
    \mathbb{I}\otimes x = x \otimes \mathbb{I} = x,\quad    
    \epsilon \otimes \epsilon = \mathbb{I},\quad
    \epsilon \otimes \sigma = \sigma \otimes \epsilon = \sigma,\quad
    \sigma \otimes \sigma = \mathbb{I} \oplus \epsilon\,,
\end{align}
with $x=\{\mathbb{I},\epsilon,\sigma\}$.
These fusion algebras can be derived from $\mathbb{Z}_2$ gauging of $\mathbb{Z}_M$ symmetries \cite{Kobayashi:2024cvp,Kobayashi:2024yqq}.
Note that the Ising algebra has the $\mathbb{Z}_2$ symmetry, where $\sigma$ transforms as $\sigma \to - \sigma$ while $\mathbb{I}$ and $\epsilon$ remain unchanged.
See Appendix \ref{app} for details.
We will also use the $\mathbb{Z}_3$ Tambara-Yamagami fusion algebra, which will be shown explicitly later.

\section{Phenomenological features of non-invertible selection rules}
\label{sec:features}
Here, we explain three specific features of non-invertible selection rules from the phenomenological point of view.

\subsection{Radiative vertices}
\label{sub2.1}
The first nature of the non-invertible selection rule is that it can be broken at the loop level even though it is preserved in the Lagrangian at the tree level \cite{Kaidi:2024wio,Heckman:2024obe,Funakoshi:2024uvy}. Note here that we discuss only cases of finite loop-diagrams which can be applicable to phenomenology.

\subsubsection{Three point vertices with trilinear terms}
\label{subsub301}
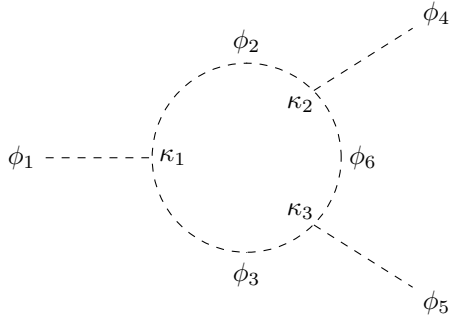
\begin{figure}[H]
\centering
\begin{tikzpicture}[
    every node/.style={font=\small},
    >=latex
]
  \node[ellipse, draw, dashed,
        minimum width=2.5cm, minimum height=2.5cm] (cluster) {};
  \node (kappa1) at (cluster.west) [xshift=8pt] {$\kappa_1$};
  \node (kappa2) at (cluster.north east) [xshift=-5pt,yshift=-5pt] {$\kappa_2$};  
  \node (kappa3) at (cluster.south east) [xshift=-5pt,yshift=5pt] {$\kappa_3$};  

  % Left
  \node (phi1) at ($(kappa1)+(-2cm,0cm)$) {$\phi_1$};
  \draw[dashed] (phi1) -- (kappa1);

  \node (phi2) at (cluster.north) [yshift=8pt] {$\phi_2$};
  \node (phi3) at (cluster.south) [yshift=-8pt] {$\phi_3$};
  \node (phi6) at (cluster.east) [xshift=8pt] {$\phi_6$};
  \draw[dashed] (phi2) -- (cluster.north);
  \draw[dashed] (phi3) -- (cluster.south);

  % Right
  \node (phi4) at ($(kappa2)+(1.8cm,1.2cm)$) {$\phi_4$};
  \node (phi5) at ($(kappa3)+(1.8cm,-1.2cm)$) {$\phi_5$};
  \draw[dashed] ($(kappa2)-(-5pt,-5pt)$) -- (phi4);
  \draw[dashed] ($(kappa3)-(-5pt,5pt)$) -- (phi5);
\end{tikzpicture}
\caption{Diagram of three point vertices with trilinear terms.
Note here that $\phi_{1-6}$ can be replaced by fermions unless the vertex-mass dimension exceeds four.}
\label{fig:212}
\end{figure}

Let us consider the three point vertices at one-loop level in Fig.~\ref{fig:212}, which can be constructed by 
three three-point terms as follows:
\begin{align}
\kappa_1 \phi_1 \phi_2\phi_3
+\kappa_2 \phi_2 \phi_4\phi_6
+\kappa_3 \phi_3 \phi_5\phi_6,
\label{eq.212}
\end{align}
where $\kappa_{1,2,3}$ has a mass dimension when $\phi_{1-6}$ are bosons, while  $\kappa_{1,2,3}$ has a dimensionless when two of the three particles are fermions; e.g., $\phi_1$ is a boson, and  $\phi_2,\ \phi_3$ are fermions. 
We can show generic fusion algebraic relations, which allow the coupling terms in Eq.~(\ref{eq.212}) at tree level, and forbid $\phi_1 \phi_4 \phi_5$ at tree level, but induce $\phi_1 \phi_4 \phi_5$ at one loop level.
For example, in the Fibonacci algebra, one simply assigns $\tau$ to $\phi_{1,2,3,6}$
and $\mathbb{I}$ to $\phi_{4,5}$.

\subsubsection{Four point vertices with trilinear terms}
\label{subsub302}
\begin{figure}[H]
\centering
\begin{tikzpicture}%[
    %node distance=1cm,
 %   every node/.style={font=\small},
 %   >=latex
%]
  %
  \node[ellipse, draw, dashed,
        minimum width=2.5cm, minimum height=2.5cm] (cluster) {};
  \node (kappa1) at (cluster.north west) [xshift=5pt,yshift=-5pt] {$\kappa_1$};
  \node (kappa2) at (cluster.south west) [xshift=5pt,yshift=5pt] {$\kappa_2$};
  \node (kappa3) at (cluster.north east) [xshift=-5pt,yshift=-5pt] {$\kappa_3$};  
  \node (kappa4) at (cluster.south east) [xshift=-5pt,yshift=5pt] {$\kappa_4$};  

  % Left
  \node (phi1) at ($(kappa1)+(-1.8cm,1.2cm)$) {$\phi_1$};
  \node (phi2) at ($(kappa2)+(-1.8cm,-1.2cm)$) {$\phi_2$};
  \draw[dashed] ($(kappa1)-(5pt,-5pt)$) -- (phi1);
  \draw[dashed] ($(kappa2)-(5pt,5pt)$) -- (phi2);

  \node (phi3) at (cluster.north) [yshift=8pt] {$\phi_3$};
  \node (phi4) at (cluster.south) [yshift=-8pt] {$\phi_4$};
  \node (phi7) at (cluster.west) [xshift=-8pt] {$\phi_7$};
  \node (phi8) at (cluster.east) [xshift=8pt] {$\phi_8$};
  \draw[dashed] (phi3) -- (cluster.north);
  \draw[dashed] (phi4) -- (cluster.south);

  % Light
  \node (phi5) at ($(kappa3)+(1.8cm,1.2cm)$) {$\phi_5$};
  \node (phi6) at ($(kappa4)+(1.8cm,-1.2cm)$) {$\phi_6$};
  \draw[dashed] ($(kappa3)-(-5pt,-5pt)$) -- (phi5);
  \draw[dashed] ($(kappa4)-(-5pt,5pt)$) -- (phi6);
\end{tikzpicture}
\caption{Diagram of four point vertices with trilinear terms.
Note here that $\phi_{1-8}$ can be replaced by fermions unless the vertex-mass dimension exceeds four.}
\label{fig:213}
\end{figure}
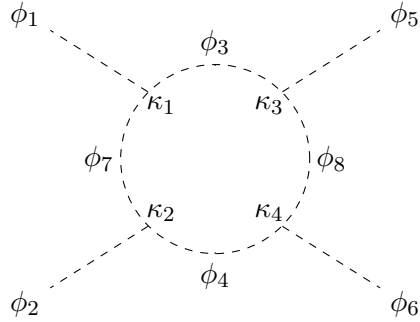

Let us consider the four point vertices at one-loop level in Fig.~\ref{fig:213}, which can be constructed by 
four three-point terms as follows:
\begin{align}
\kappa_1 \phi_1 \phi_3\phi_7
+\kappa_2 \phi_2 \phi_4\phi_7
+\kappa_3 \phi_3 \phi_5\phi_8
+\kappa_4 \phi_4 \phi_6\phi_8,
\label{eq.213}
\end{align}
where all the nature of $\kappa_{1,2,3,4}$ are the same as the case of Eq.~(\ref{eq.212}). 
Similar to section \ref{subsub301},  we can show generic fusion algebraic relations, which allow the coupling terms in Eq.~(\ref{eq.213}) at tree level, and forbid $\phi_1 \phi_2 \phi_5 \phi_6$ at tree level, but induce $\phi_1 \phi_2 \phi_5 \phi_6$ at one loop level.
For example, in the Fibonacci algebra, one assigns $\tau$ to $\phi_{1,3,4,7,8}$
and $\mathbb{I}$ to $\phi_{2,5,6}$.

\subsubsection{Four point vertices with two trilinear and one quartic terms}
\label{subsub303}
\begin{figure}[H]
\centering
\begin{tikzpicture}%[
    %node distance=1cm,
 %   every node/.style={font=\small},
 %   >=latex
%]
  %
  \node[ellipse, draw, dashed,
        minimum width=2.5cm, minimum height=2.5cm] (cluster) {};
  \node (kappa1) at (cluster.north west) [xshift=5pt,yshift=-5pt] {$\kappa_1$};
  \node (kappa2) at (cluster.south west) [xshift=5pt,yshift=5pt] {$\kappa_2$};
  \node (lambda) at (cluster.east) [xshift=-8pt] {$\lambda$};  
    % Left
  \node (phi1) at ($(kappa1)+(-1.8cm,1.2cm)$) {$\phi_1$};
  \node (phi2) at ($(kappa2)+(-1.8cm,-1.2cm)$) {$\phi_2$};
  \draw[dashed] ($(kappa1)-(5pt,-5pt)$) -- (phi1);
  \draw[dashed] ($(kappa2)-(5pt,5pt)$) -- (phi2);
  \node (phi4) at (cluster.north) [yshift=8pt] {$\phi_4$};
  \node (phi5) at (cluster.south) [yshift=-8pt] {$\phi_5$};
  \node (phi3) at (cluster.west) [xshift=-8pt] {$\phi_3$};
  \draw[dashed] (phi4) -- (cluster.north);
  \draw[dashed] (phi5) -- (cluster.south);
  % Light
  \node (phi6) at ($(lambda)+(1.8cm,1.2cm)$) {$\phi_6$};
  \node (phi7) at ($(lambda)+(1.8cm,-1.2cm)$) {$\phi_7$};
  \draw[dashed] ($(lambda)+(8pt,0cm)$) -- (phi6);
  \draw[dashed] ($(lambda)+(8pt,0cm)$) -- (phi7);
\end{tikzpicture}
\caption{Diagram of four point vertices with trilinear and quartic terms.
Note here that $\phi_{1-3}$ can be replaced by fermions unless the vertex-mass dimension exceeds four.}
\label{fig:215}
\end{figure}
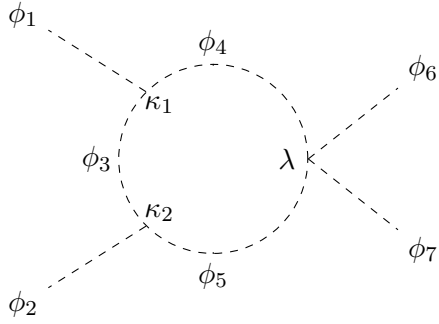

Let us consider the four point vertices at one-loop level in Fig.~\ref{fig:215}, which can be constructed by 
two trilinear and one quartic terms:
\begin{align}
\kappa_1 \phi_1 \phi_3\phi_4
+\kappa_2 \phi_2 \phi_3\phi_5
+\lambda \phi_4 \phi_5\phi_6\phi_7,
\label{eq.215}
\end{align}
where all the nature of $\kappa_{1,2}$ are the same as the case of Eq.~(\ref{eq.212}).
%%%
Similar to section \ref{subsub301},  we can show generic fusion algebraic relations, which allow the coupling terms in Eq.~(\ref{eq.215}) at tree level, and forbid $\phi_1 \phi_2 \phi_6 \phi_7$ at tree level, but induce $\phi_1 \phi_2 \phi_6 \phi_7$ at one loop level.
For example, in the Fibonacci algebra, one assigns $\tau$ to $\phi_{2,3,4,5}$
and $\mathbb{I}$ to $\phi_{1,6,7}$.

 All of loop diagrams shown in Figs.~\ref{fig:212}-
 \ref{fig:215} would be important in model building and they would lead to various interesting phenomenological aspects.
 For illustration, we will show some examples of models inducing loop diagrams of Fig.~\ref{fig:212} 
 in the next sections.
 Other loop diagrams would lead to interesting models.
We would study them elsewhere.
We may have other loop diagrams to break selection rules at the tree level including higher loops.

\subsection{A remnant symmetry}
\label{sec:remnant}
Even though there exists no mechanism in subsection \ref{sub2.1}, the non-invertible selection rules can play a role in realizing radiative seesaw models.
Here, we show a typical example of Ma model~\cite{Ma:2006km}.
For simplicity, we apply the simple non-invertible selection rule realized by the Ising 
fusion algebra \eqref{eq:Ising} (instead of $\mathbb{Z}_2$ symmetry).

We list the particle contents and their charge-assignments in Table~\ref{tab:2.2}, where we also show the $\mathbb{Z}_2$ assignments in the original Ma-model. 
New fields $N_R$ and $\eta$ are assigned to $\sigma$. 
One finds that the nontrivial representation $\sigma$ corresponds to $\mathbb{Z}_2$ odd.
Under the selection rule, the relevant terms are as follows:
%%%
\begin{align}
& 
y_\ell\overline{L_{L}}\ell_{R}H
+y_\eta\overline{L_{L}}N_{R}(i\sigma_2) \eta^*
+M_N \overline{N_R} N_R^C +{\rm h.c.}\\
%%%
-&\mu_H^2|H|^2
-\mu_\eta^2|\eta|^2
+\lambda_1 |H|^4+\lambda_2 |\eta|^4 +\lambda_3 |H|^2|\eta|^2
+\lambda_4 |H^\dagger\eta|^2+\lambda_5[(H^\dagger\eta)^2+{\rm c.c.}].
\end{align}
%%%
One finds that the above terms are the same as the ones of the Ma model.
Thus, the Ising fusion rule of this scenario is nothing but an equivalent of the $\mathbb{Z}_2$ symmetry to realize the radiative seesaw model,
and this symmetry never be broken unless one extends Ma model to something new scenarios. 

Note that the existence of $\mathbb{Z}_2$ symmetry originates from the fact that the fusion algebra is invariant under the $\mathbb{Z}_2$ symmetry $\sigma \rightarrow -\sigma$, and odd number of external $\sigma$ particles are forbidden at any loop order \cite{Kaidi:2024wio}. 
Furthermore, when we replace $\sigma$ in the Ising fusion rule with $\tau$ in the Fibonacci fusion rule, we arrive at the same conclusion; the effective $\mathbb{Z}_2$ symmetry still remains by a combination with gauge charge and spin conservations.

\begin{table}%[tbc]
    \centering
\begin{tabular}{|c||c|c|c||c|c|}\hline\hline  
& ~$\overline{ L_L}$~ & ~$ \ell_R$~ & ~$N_R$ ~&~ {$H$} ~&~ {$\eta$}~\\\hline\hline 
%%%
$SU(2)_L$   & $\bm{2}$  & $\bm{1}$  & $\bm{1}$  & $\bm{2}$& $\bm{2}$      \\\hline 
$U(1)_Y$    & $\frac12$  & $-1$ & $0$ & {$\frac12$} & {$\frac12$}      \\\hline
${\rm Ising\ Rule}\, (\mathbb{Z}_2)$   & $\mathbb{I}\ (+)$  & $\mathbb{I}\ (+)$ & $\sigma\ (-)$ & $\mathbb{I}\ (+)$& $\sigma\ (-)$  \\\hline 
\end{tabular}
\caption{Charge assignments of the relevant fields
under $SU(2)_L\otimes U(1)_Y \otimes {\rm Ising \ Rule}\ (\mathbb{Z}_2)$, where the notation of this rule is shown in section \ref{sec:selection-rule}.}
\label{tab:2.2}
\end{table}

\subsection{Texture realizations for mass matrix via non-invertible selection rules}
\label{sec:textures}
The last one is to realize a predictive texture through a non-invertible selection rule.
Some precursors have realized predictive textures for quark sector~\cite{Kobayashi:2024cvp,Kobayashi:2025znw} and lepton sector~\cite{Kobayashi:2025ldi} under the Weinberg operators in the framework of models with non-invertible selection rules. 
In particular, they start from $D_M \cong \mathbb{Z}_N\rtimes \mathbb{Z}_2$ symmetry. (For more details, see Appendix \ref{app}.) 
For the fusion algebras realized by the irreducible representation of $D_M$, we restrict ourselves to the $\mathbb{Z}_2$ invariant modes, whose fusion algebras include the Fibonacci 
fusion algebra \eqref{eq:Fibo} for $M=3$ and Ising fusion algebra \eqref{eq:Ising} for $M=4$. 
Then, it leads to interesting Yukawa textures, including the nearest neighbor interaction texture and its extension in both quark and lepton sectors. 
The fermion masses and mixing angles as well as the CP phases are successfully explained. 
Furthermore, the other Yukawa texture can address the strong CP problem without introducing axions by assuming the CP invariance in UV theory. 
Such a $\mathbb{Z}_2$ gauging is motivated by the selection rules of chiral zero modes in string compactifications such as type II intersecting/magnetized D-branes \cite{Kobayashi:2024yqq,Funakoshi:2024uvy}. Other selection rules are recently found in Ref. \cite{Dong:2025pah} in the context of heterotic string theory.
%%%

In our paper, we will show a construction of predictive textures for the lepton sector under the framework of Ma model below.
In considering a texture for a radiative-induced mass matrix,
particles which run in the loop could contribute to the structure of the texture. That leads us to a little difficult situation to realize a predictive texture. 
In most of cases, fermions inside the loop contribute to the structure.~\footnote{Bosons inside the loop also contribute to the texture in Ref.~\cite{Hehn:2012kz}.}
In order to avoid such a difficulty,
we might need to impose some specific hierarchies among particles inside the loop
or
diagonal mass matrix for the fermions inside the loop.
%%%
However, any non-invertible selection rules have the latter feature that Majorana fermion mass matrix is diagonal when a non-trivial representation is assigned \cite{Kobayashi:2025ldi}, that is,
\begin{align}
\sigma\otimes\sigma \supset \mathbb{I},
\end{align}
where $\sigma\neq 1$ is arbitrary representation.
This is because its difficulty is automatically evaded by this symmetry if one uses
Majorana neutral fermions inside the loop as in the case of Ma model.
%%%
Below, we will see a concrete model in the latter case.

%%%%%%%%%%%%%%%%%%%%%%%%%%%%%%%%%%%%%%%%%%%%%%%%%%%%%%%%%%%%%%%%%
\section{Examples for radiative vertices}
\label{sec:examples}
Here, we show some model-examples to realize radiative finite vertices in subsection \ref{sub2.1}.

\subsection{Model 1}\label{subsec_m1}
The first example is to apply for the type-II neutrino model that requires an $SU(2)_L$ triplet boson with $U(1)_Y=1$
where we denote the boson as $\Delta$.
The model induces the Majorana neutrino masses via a vacuum expectation value of this boson $v_\Delta$.  
That implies that $v_\Delta$ has to be tiny.
The model can theoretically explain the smallness of $v_\Delta$ by generating at finite one-loop level.
The original scenario is found in Ref.~\cite{Kanemura:2012rj}.
%%% %%%
$v_\Delta$ is given in terms of some parameters of the Higgs potential via solving the tad-pole conditions and it is mainly proportional to the one mass dimensional parameter $\mu_\Delta$ that comes from $H^T (i\sigma_2) \Delta^\dag H$ where $\sigma_2$ is the second Pauli matrix.
Hence, we obtain
\begin{align}
m_\nu\propto v_\Delta \propto \mu_\Delta.
\end{align}
In our model, however, the term $\mu_\Delta$ is not allowed by a non-invertible selection rule at tree level, but allowed at loop level, as we will show below.
%%% %%%
%\if0
At first, we assign $\phi_{1,2,3,4,5,6}$ to be $\phi_1\sim({\bf3},1)$, $\phi_{2,3}\sim({\bf2},1/2)$, $\phi_{4,5}\sim({\bf2},1/2)$, and $\phi_6\sim({\bf1},1)$ under $SU(2)_L\otimes U(1)_Y$.
%\fi
%%%
We change these notations as $\Delta\equiv \phi_1$, $\eta\equiv \phi_{2,3}$, $H\equiv \phi_{4,5}$, and $\phi_6\equiv S$ for our convenience.
Here, $\eta$ is an inert doublet boson, $S$ is an inert singlet boson, and $\varphi$ is a singlet boson with nonzero VEV $\langle\varphi\rangle\equiv v'/\sqrt2$ that spontaneously violates $U(1)_{B-L}$ symmetry.
$H$ is identified as the SM Higgs boson.
Then, we apply the Ising fusion rule
%%%
\footnote{This is the minimum realization. 
If one applies the Fibonacci fusion rule, one may have the following two additional terms $\bar L_L \eta^* \ell_R$ and $H^\dag \Delta \eta$.
And the second term would violate the inert feature of $\eta$.} 
%%%
 and a $U(1)_{B-L}$ symmetry\footnote{The Three right-handed neutrinos are needed to cancel the chiral anomalies. But, we neglect these particles because it is out of our interest.} for the relevant fields and we list their assignments in Table~\ref{tab:2}.
 The $U(1)_{B-L}$ symmetry plays a crucial role in forbidding the term $(H^\dag\eta)^2$ that spoils our model due to generating the logarithmic divergence.
%
%%%%%%%%%%%%%%%%%%%%%%%%%%%%%%%%%%%%%%%%%%%%%%%%%%%%%%%%%%%%%%%%%
\begin{table}[H]%[tbc]
    \centering
\begin{tabular}{|c||c|c||c|c|c|c|c|}\hline\hline  
& ~$\overline{ L_L}$~ & ~$ \ell_R$~ & ~$ H$ ~&~ {$\eta$} ~&~ {$\Delta$}
~&~ {$S$} ~&~ {$\varphi$}~  \\\hline\hline 
%%%
$SU(2)_L$   & $\bm{2}$  & $\bm{1}$  & $\bm{2}$  & $\bm{2}$& $\bm{3}$& $\bm{1}$& $\bm{1}$     \\\hline 
$U(1)_Y$    & $\frac12$  & $-1$ & $\frac12$ & {$\frac12$} & {$+1$}& {$0$} & {$0$}    \\\hline
${\rm Ising\ Rule}$   & $\sigma$  & $ \sigma$  & $\mathbb{I}$ & $\sigma$& $\epsilon$
& $\tau$& $\mathbb{I}$\\\hline 
%%%
$U(1)_{B-L}$   & $-\ell$  & $\ell$  & $0$ & $\ell$& $2\ell$
& $\ell$ & $-2\ell$\\\hline 
\end{tabular}
\caption{Charge assignments of the relevant fields
under $SU(2)_L\otimes U(1)_Y \otimes {\rm Ising\ Rule}\otimes {\rm Lepton\ Number}$.}
\label{tab:2}
\end{table}

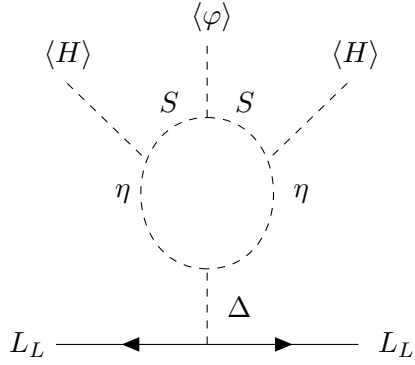
\begin{figure}[H]
    \centering
\begin{tikzpicture}
  \begin{feynman}
    \vertex[label=left:\(\ L_L\)]  (a) at (0,0);
    \vertex[label=right:\(\ L_L\)] (b) at (4,0); 
    \vertex (c) at (2,0);
    \vertex (d) at (2,1);
    \vertex (e) at (2,3);
    \vertex[label=above:\(\ \langle H\rangle\)] (f) at (0.1,3.5);
    \vertex[label=above:\(\ \langle H\rangle\)] (g) at (3.9,3.5); 
    \vertex (h) at (0.866+2, 0.5+2);
    \vertex (i) at (2-0.866, 0.5+2);
    \vertex[label=above:\(\ \langle \varphi\rangle\)] (j) at (2,4); 
    \diagram*{
      (c) -- [fermion] (a),
      (c) -- [fermion] (b),
      (c) -- [scalar,edge label'=\(\ \Delta\)] (d),
      (d) -- [scalar, half left, looseness=1.5,edge label=\(\ \eta\)] (e),
      (e) -- [scalar, half left, looseness=1.5,edge label=\(\ \eta\)] (d),
      (h) -- [scalar] (g),
      (i) -- [scalar] (f),
      (e) -- [scalar] (j)
      };
    \node at (1.5,3.2) {\(S\)};
    \node at (2.5,3.2) {\(S\)};
  \end{feynman}
\end{tikzpicture}
    \caption{Neutrino masses in the type-II neutrino model at finite one-loop level.}
    \label{fig:typeII}
\end{figure}

The valid Lagrangian is given by
\begin{align}
-{\cal L}\sim y_\ell \overline{L_{L}} H \ell_R + y_\nu \overline{L_L} (i\sigma_2)\Delta^\dagger L_L^c 
+{\rm h.c.},
\end{align}
where the first and second terms respectively induce the masses of charged leptons and the masses of active neutrinos. This is nothing but the normal type-II seesaw.
The Higgs potential at the tree level is found as
\begin{align}
{\cal V}  
&= -\mu_H^2 |H|^2-\mu_\eta^2 |\eta|^2-\mu_\Delta^2 {\rm Tr}[\Delta^\dagger \Delta] + (\rm Trivial\ quartic\ terms)\nn\\
&+\mu \eta^T \Delta^\dag \eta + \mu(H^\dag\eta)S+ \mu' S^2\varphi .
\end{align}
The second line of the above equation provides $H^T (i\sigma_2) \Delta^\dag H$ at the one-loop level in Fig.~\ref{fig:typeII}, although this term is not allowed at tree level by the Ising fusion algebra.

It would be worthwhile to mention that $\eta$ can be a DM candidate and its stability is assured by the Ising fusion rule. Then, the neutrino masses are generated at one-loop level, as shown in Fig.~\ref{fig:typeII}. 
Compared our model with the original model in Ref.~\cite{Kanemura:2012rj}, simpler model building has been achieved. 
The original one introduces two additional symmetries ($U(1)_L$, $\mathbb{Z}_2$) with four additional bosons ($s^0_1$,$s^0_2$,$\eta$, $\Delta$), but our model requests one additional selection rule (Ising fusion rule) with two additional bosons ($\Delta,\ \eta$).

\subsection{Model 2}
Another example is to construct an extended Zee-Babu type neutrino mass model~\cite{Zee:1985id, Babu:1988ki} applying for $\mathbb{Z}_3$ Tambara-Yamagami (TY) fusion rule which is one of the non-invertible selection rules~\cite{Chang:2018iay}, i.e.,
\begin{align}
\label{eq:TY}
    \eta^3 = \mathbb{I},\quad
    \eta \otimes {\cal N} = {\cal N},\quad
    {\cal N}\otimes {\cal N} = \mathbb{I} + \eta + \eta^2.
\end{align}
For more details, see Appendix \ref{app}. 
The neutrino model provides the neutrino mass matrix at two-loop level 
as a leading contribution by introducing a singly-charged boson $S_1^-$ and a doubly-charged boson $k^{++}$. 
%%%
The bare Lagrangian is given by
\begin{align}
-{\cal L}\sim y_\ell\overline{L_{L}} H \ell_R + f \overline{L_L} (i\sigma_2) L_L^c S_1^- + g \overline{\ell_R^C} \ell_R k^{++}
+{\rm h.c.},\label{eq:zeebabu}
\end{align}
where the first term induces the masses of charged-leptons, and the second and third terms contribute to the masses of active neutrinos. 
Then, the active neutrino mass matrix is proportional to $\mu_{kss}$ where it is a coefficient of $k^{++}S_1^-S_1^-$:
\begin{align}
m_\nu\propto  \mu_{kss}.
\end{align}
In our extended model, the term corresponding to $\mu_{kss}$ is induced at the one-loop level as the leading order. 
In addition to the original Zee-Babu model, we introduce a singly-charged boson $S_2^+$ 
and a singlet neutral boson $S_0$.
In this case, the identifications are as follows 
 $\phi_{1},\phi_{4,5},\phi_{2,3},\phi_{6}$ to be $({\bf1},+2)$, $({\bf1},-1)$, $({\bf1},+1)$, $({\bf1},0)$ under $SU(2)_L\otimes U(1)_Y$.
These fields correspond to $k^{++}\equiv \phi_1$, $S^-_1\equiv \phi_{4,5}$, $S^+_2\equiv \phi_{2,3}$, $S_0\equiv \phi_{6}$. 
Therefore, the relevant terms are given by
 \begin{align}
\mu^2_S S_0^2 + \mu_0  k^{++} S^-_2 S^-_2 + \mu_1 S^+_2 S^-_1 S_0 + {\rm h.c.}. \label{eq:model2}
\end{align}
Following subsection~\ref{subsub301}, the $\mu_{kss}$ is written by
 \begin{align}
\mu_{kss} \sim \frac{\mu_0 \mu_1^2}{(4\pi)^2 m_S^2},
\end{align}
where $m_S$ is the largest mass among $S_0,\ S^-_1,\ S^+_2$.
%%%
Then, we apply the $\mathbb{Z}_3$ TY fusion rule \eqref{eq:TY} for the relevant fields, and we list their assignments in Table~\ref{tab:3}.

\begin{table}[H]%[tbc]
    \centering
\begin{tabular}{|c||c|c||c|c|c|c|c|}\hline\hline  
& ~$\overline{ L_L}$~ & ~$ \ell_R$~ & ~$ H$ ~&~ {$k^{++}$} ~&~ {$S^-_1$}~ & ~ {$S^+_2$}~
& ~ {$S_0$}~\\\hline\hline 
%%%
$SU(2)_L$   & $\bm{2}$  & $\bm{1}$  & $\bm{2}$  & $\bm{1}$& $\bm{1}$ & $\bm{1}$& $\bm{1}$     \\\hline 
$U(1)_Y$    & $\frac12$  & $-1$ & $\frac12$ & $2$ & {$-1$} & {$+1$} & {$0$}    \\\hline
${\rm TY}$   & $\cal N$  & $\cal N$  & $\mathbb{I}$ & $\eta$& $\eta$ & ${\cal N}$
& ${\cal N}$ \\\hline 
\end{tabular}
\caption{Charge assignments of the relevant fields
under $SU(2)_L\otimes U(1)_Y \otimes {\rm TY}$.}
\label{tab:3}
\end{table}

\begin{figure}[H]
    \centering
\begin{tikzpicture}
  \begin{feynman}
   \draw[dashed] (5,3) ellipse [x radius=2cm, y radius=1cm];    %
    \vertex[label=left:\(\ L_L\)]  (a) at (0,0); 
    \vertex[label=right:\(\ L_L\)] (b) at (10,0); 
    \vertex (c) at (5,0);
    \vertex (d) at (5,2);
    \vertex (e) at (5,4);
    \vertex (f) at (2,0);
    \vertex (g) at (8,0); 
    \vertex (h) at (3.5,0); 
    \vertex (i) at (6.5,0); 
    \vertex[label=below:\(\ \langle H \rangle\)] (j) at (3.5,-1); 
    \vertex[label=below:\(\ \langle H \rangle\)] (k) at (6.5,-1); 
    \vertex (l) at (3.5,2.35); 
    \vertex (m) at (6.5,2.35); 
    \vertex[label=below:\(\ L_L\)] at (2.75,0);
    \vertex[label=below:\(\ l_R\)] at (4.25,0);
    \vertex[label=below:\(\ l_R\)] at (5.75,0);
    \vertex[label=below:\(\ L_L\)] at (7.25,0);
    \node at (2.3,1.1) {\( S_1^-\)};
    \node at (7.7,1.1) {\( S_1^-\)};
    \node at (4,1.8) {\( S_2^-\)};
    \node at (6,1.8) {\( S_2^-\)};
    \node at (5,4.3) {\( S_0\)};
    \diagram*{
      (f) -- [fermion] (a),
      (f) -- [fermion] (h),
      (h) -- [fermion] (c),
      (g) -- [fermion] (b),
      (g) -- [fermion] (i),
      (i) -- [fermion] (c),
      (c) -- [scalar,edge label'=\(\ k^{--}\)] (d),
%      (d) -- [scalar, half left, looseness=1.5] (e),
%      (e) -- [scalar, half left, looseness=1.5] (d),
      (f) -- [scalar] (l), 
      (g) -- [scalar] (m), 
      (h) -- [scalar] (j),
      (i) -- [scalar] (k),
    };
  \end{feynman}
\end{tikzpicture}
    \caption{Neutrino masses in an extended Zee-Babu model at finite three-loop level.}
    \label{fig:ZeeBabu}
\end{figure}
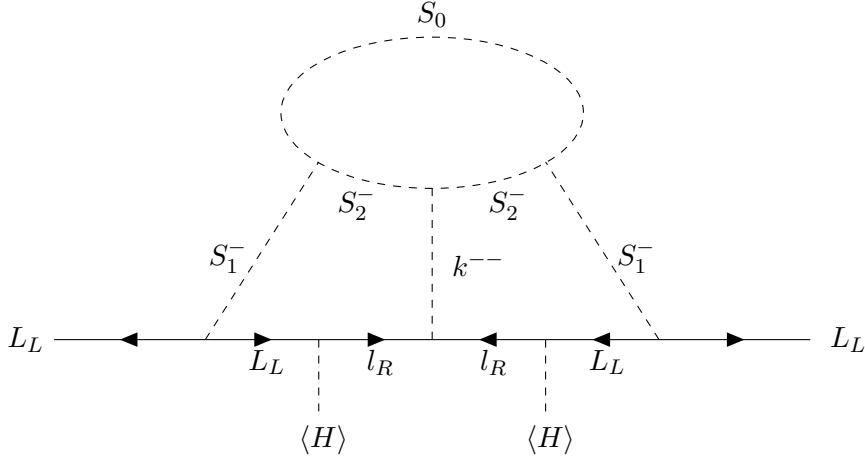

Under these assignments, $k^{++}S_1^-S_1^-$ as well as  $\overline{L_L} (i\sigma_2) L_L^c S_2^-$ and $S_2^+S_1^-$ are forbidden but Eq.~(\ref{eq:model2}) is allowed. 
We have achieved an extended Zee-Babu model at three-loop level, as shown in Fig.~\ref{fig:ZeeBabu}.
%%%
%%%

%%%%%%%%%%%%%%%%%%%%%%%%%%%%%%%%%%%%%%%%%%%%%%%%%%%%%%%%%%%%%%%%%
\section{Examples for lepton mass textures in Ma model}
\label{sec:Matextures}
\begin{figure}
    \centering
\begin{tikzpicture}
  \begin{feynman}
    \vertex[label=left:\(\ L_L\)] (a) at (0,0);       
    \vertex[label=right:\(\ L_L\)] (b) at (6,0);       
    \vertex (c) at (1,0);       
    \vertex[crossed dot, label=above:\(\ N_R\)] (m) at (3,0);
    \vertex[label=below:\(\ \times\), yshift=7.5pt] at (3,0);       
    \vertex (d) at (5,0);       
    \vertex (e) at (3,2.35);       
    \vertex[label=above:\(\ \langle H\rangle\)] (f) at (1.5,3.5);     %  
    \vertex[label=above:\(\ \langle H\rangle\)] (g) at (4.5,3.5);
    \vertex[label=left:\(\ \eta\)] (h) at (1.4,1.6);       
    \vertex[label=right:\(\ \eta\)] (i) at (4.5,1.6);       
    \diagram* {
      (c) -- [fermion] (a),    
      (m) -- [fermion] (c),    
      (m) -- [fermion] (d),    
      (d) -- [fermion] (b),                         
      (c) -- [scalar, half left, looseness=2] (d),  
      (f) -- [scalar] (e),                         
      (e) -- [scalar] (g),                         
    };
  \end{feynman}
\end{tikzpicture}
    \caption{Neutrino masses in Ma model with the Ising fusion rule.}
    \label{fig:Ma}
\end{figure}
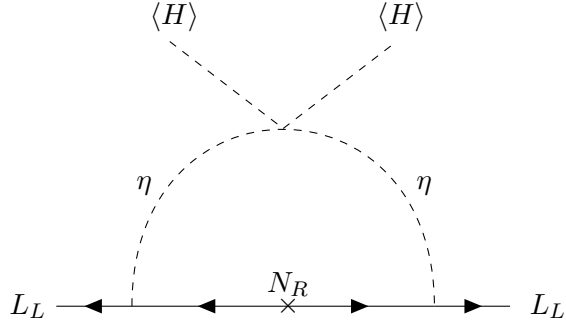

We show some predictive textures in Ma model applying the Ising fusion rule.
Here, we introduce only two right-handed neutral fermions $N_{R_{1,2}}$ and assign them 
to $\epsilon,\sigma$. The inert boson $\eta\equiv[\eta^+,(\eta_R+i \eta_I)/\sqrt2]^T$ is assigned to $\sigma$.
The left-handed and right-handed leptons; $\overline{L_L}\equiv (\overline{L_{L_e}},\overline{L_{L_\mu}},\overline{L_{L_\tau}})$, $\ell_R\equiv(e_R,\mu_R,\tau_R)$ are uniformly assigned to
$(1,\epsilon,\sigma)$ or $(\sigma,\epsilon,1)$, where the SM Higgs boson is the identity $\mathbb{I}$ under the non-invertible selection rule and $\mathbb{Z}_2$ symmetry is implicitly imposed in order to forbid Dirac Yukawa couplings such as $\overline{L_{L_\mu}}N_{R_1}(i\sigma_2) H^*$ and $\overline{L_{L_\tau}}N_{R_2}(i\sigma_2) H^*$. The relevant particles and their assignments are summarized in Table~\ref{tab:4}.
\begin{table}[H]%[tbc]
    \centering
\begin{tabular}{|c||c|c|c||c|c|}\hline\hline  
& ~$\overline{ L_L}$~ & ~$ \ell_R$~ & ~$N_R$ ~&~ {$H$} ~&~ {$\eta$}~\\\hline\hline 
%%%
$SU(2)_L$   & $\bm{2}$  & $\bm{1}$  & $\bm{1}$  & $\bm{2}$& $\bm{2}$      \\\hline 
$U(1)_Y$    & $\frac12$  & $-1$ & $0$ & {$\frac12$} & {$\frac12$}      \\\hline
${\rm Ising\ Rule}$   & $(\mathbb{I},\epsilon,\sigma)$  & $(\mathbb{I},\epsilon,\sigma)$  & $(\epsilon,\sigma)$ & $\mathbb{I}$& $\sigma$        \\\hline 
\end{tabular}
\caption{Charge assignments of the relevant fields
under $SU(2)_L\otimes U(1)_Y \otimes {Ising \ Rule}$ where the notation of this rule is the same as the one in subsection~\ref{subsec_m1}.
}\label{tab:4}
\end{table}
The terms $\lambda_5(H^\dag \eta)^2+$h.c. are allowed by these symmetries and the charged-lepton Yukawa mass matrix and the right-handed neutral fermion mass matrix are diagonal; $\sum_{l=e,\mu,\tau} y_{l} \overline{L_{L_l}} \ell_{R_l} H$
and $\sum_{a=1,2} M_a \overline{N_{R_a}^C} N_{R_a}$.
Thus, the structure of the active neutrino mass matrix mainly arises from the following terms:
\begin{align}
& 
y_1\overline{L_{L_e}}N_{R_2}(i\sigma_2) \eta^*
+
y_4\overline{L_{L_\tau}}N_{R_1}(i\sigma_2) \eta^*
+
\overline{L_{L_e}}\left(y_2 N_{R_1}+y_3 N_{R_2}\right)(i\sigma_2) \eta^*
+{\rm h.c.}\nn\\
&=
\begin{pmatrix}
\overline{L_{L_e}}  \\ 
\overline{L_{L_\mu}}  \\ 
\overline{L_{L_\tau}}  \\ 
\end{pmatrix}^T
\begin{pmatrix}
0 & y_1 \\ 
y_2 & y_3 \\ 
y_4 & 0 \\ 
\end{pmatrix}
\begin{pmatrix}
N_{R_1} \\ 
N_{R_2}  \\ 
\end{pmatrix} (i\sigma_2) \eta^*
\equiv \overline{L_L} y_D N_R (i\sigma_2) \eta^*
.\label{massmat}
\end{align}
Then the neutrino mass matrix is determined by the one-loop diagram in \figref{fig:Ma}
\begin{align}
(m_\nu)_{ij}
&=\frac{1}{(4\pi)^2}\sum_{a=1}^2 (y_D)_{ia} M_a (y_D^T)_{aj}
\left(\frac{m_R^2}{M^2_a-m_R^2}\ln\left[\frac{m_R^2}{M_a^2}\right]
-
\frac{m_I^2}{M^2_a-m_I^2}\ln\left[\frac{m_I^2}{M_a^2}\right]
\right)\\
&\equiv (y_D)_{ia} D_a (y_D^T)_{aj}, 
\end{align}
where $D_a$ is defined by
\begin{align}
D_a\equiv \frac{M_a}{(4\pi)^2}
\left(\frac{m_R^2}{M^2_a-m_R^2}\ln\left[\frac{m_R^2}{M_a^2}\right]
-
\frac{m_I^2}{M^2_a-m_I^2}\ln\left[\frac{m_I^2}{M_a^2}\right]
\right).
\end{align}
Here, $m_R$ and $m_I$ are respectively the mass eigenvalues of $\eta_R$ and $\eta_I$,
and its mass difference is assured by nonzero coupling of $\lambda_5$.

It suggests that the structure of the neutrino mass matrix is determined by
$y_D$ and $D_a$ and one obtains the following texture from $y_D D_a y_D^T$:
\begin{align}
\begin{pmatrix}
\times & \times & 0 \\ 
\times & \times & \times \\ 
0 & \times & \times \\ 
\end{pmatrix}. \label{eq:typec}
\end{align}
The above one-zero texture is called type-C in Ref.~\cite{Barreiros:2018ndn},
and this texture provides some predictions satisfying the neutrino oscillation data~\cite{Esteban:2024eli}.

Then, $m_\nu$ can be diagonalized by PMNS matrix $U$, since the charged-lepton mass matrix is diagonal; $D_\nu(=(m_1,m_2,0))\equiv U^T m_\nu U$, in case of inverted hierarchy (IH)~\footnote{Note here that all the valid one-zero textures satisfy the neutrino oscillation data only in the case of IH.}. 
Therefore one obtains $m_\nu=U^* D_\nu U^\dagger$ and its right-hand side is written 
by experimental values; three mixings ($s_{12},s_{23},s_{13}$) with $s_{ij}=\sin \theta_{ij}$, ($i=1,2,3$) and two mass square differences $(\Delta m^2_{\rm sol},\Delta m^2_{\rm atm})$, and Dirac and Majorana phases. Thus, one derives 
\begin{align}
\frac{m_1}{m_2} = -\frac{U^*_{12} U^*_{32}}{U^*_{11} U^*_{31}},
\label{eq:type-c}
\end{align}
where we have used $(m_\nu)_{13}=0$ in Eq.~(\ref{eq:typec}).
In Eq.~(\ref{eq:type-c}), one finds the following relations
\begin{align}
c_\delta &= -\frac12\frac{[s^4_{12}(1+r_\nu)-c^4_{12}]c^2_{23}s^2_{23}+r_\nu s^2_{23}s^2_{12}c^2_{12}}{[s^2_{12}(1+r_\nu)+c^2_{12}]s_{12}c_{12}s_{23}c_{23}s_{13}},\\
%%%
c_\alpha &= \frac{-[s^4_{12}(1+r_\nu)+c^4_{12}]c^2_{23}s^2_{13}+(2+r_\nu) s^2_{23}s^2_{12}c^2_{12}}
{2\sqrt{1+r_\nu}(s^2_{23}+c^2_{23}s^2_{13})s^2_{12}c^2_{12}},
\label{eq:type-angles}
\end{align}
where $r_\nu\equiv\frac{\Delta m^2_{\rm sol}}{|\Delta m^2_{\rm atm}|}$ and $c_{ij}=\cos \theta_{ij}$.

Other promising possibility of texture is to assign 
$\overline{L_L}$ and $\ell_R$ are uniformly assigned to
$(1,\sigma,\epsilon)$ or $(\sigma,1,\epsilon)$, where the other assignments are the same as the type-C.
In this case, one finds the following texture;
\begin{align}
\begin{pmatrix}
\times & 0 & \times \\ 
0 & \times & \times \\ 
\times & \times & \times \\ 
\end{pmatrix}. \label{eq:typeb}
\end{align}
The above one-zero texture is called type-B in Ref.~\cite{Barreiros:2018ndn},
and one finds
\begin{align}
\frac{m_1}{m_2} = -\frac{U^*_{12} U^*_{22}}{U^*_{11} U^*_{21}},
\label{eq:type-b}
\end{align}
where we have used $(m_\nu)_{12}=0$ in Eq.~(\ref{eq:typeb}).
In Eq.~(\ref{eq:type-b}), one finds the following relations
\begin{align}
c_\delta &= \frac12\frac{[s^4_{12}(1+r_\nu)-c^4_{12}]s^2_{23}s^2_{13}+r_\nu c^2_{23}s^2_{12}c^2_{12}}{[s^2_{12}(1+r_\nu)+c^2_{12}]s_{12}c_{12}s_{23}c_{23}s_{13}},\\
%%%
c_\alpha &= \frac{-[s^4_{12}(1+r_\nu)+c^4_{12}]s^2_{23}s^2_{13}+(2+r_\nu) c^2_{23}s^2_{12}c^2_{12}}
{2\sqrt{1+r_\nu}(c^2_{23}+s^2_{23}s^2_{13})s^2_{12}c^2_{12}}.
\label{eq:type-mixings}
\end{align}

These textures can straightforwardly be applied to the other seesaw models.
For example, in assigning $H$ to $\sigma$ without $\eta$,
we can realize the above texture in the framework of the canonical seesaw scenario
that leads us to the same result as we discussed above.

%%%%%%%%%%%%%%%%%%%%%%%%%%%%%%%%%%%%%%%%%%%%%%%%%%%%%%%%%%%%%%%%%
\section{Conclusions}
\label{sec:con}
%%%%%%%%%%%%%%%%%%%%%%%%%%%%%%%%%%%%%%%%%%%%%%%%%%%%%%%%%%%%%%%%%

In this paper, we applied the non-invertible selection rules to radiative neutrino mass models. Conventionally, we impose certain selection rules coming from group-like symmetries to forbid interactions for fields and to ensure the stability of DM. 
We proposed an alternative approach to address the origin of tiny neutrino masses, i.e., non-invertible selection rules. 

By assigning the elements of fusion algebras to fields in quantum field theory, 
the non-invertible selection rules restrict the bare interaction terms. 
Since the non-invertible selection rules are violated by quantum corrections \cite{Heckman:2024obe,Kaidi:2024wio,Funakoshi:2024uvy}, they naturally provide the radiative neutrino masses which was shown in explicit models such as Ma model \cite{Ma:2006km}, type-II neutrino model \cite{Kanemura:2012rj}, and an extended Zee-Babu type neutrino model \cite{Zee:1985id,Babu:1988ki}. 
In addition to the radiative corrections, we discussed phenomenological implications of the remnant symmetry of non-invertible selection rules which holds at all-loop order. 
It turned out that the $\mathbb{Z}_2$ symmetry induced by the Fibonacci and the Ising fusion algebras respectively ensures the stability of DM in Ma model and type-II neutrino model. 
Note that in the case of type-II neutrino model, one can perform more simpler model building due to the non-invertible selection rule.

So far, fields were labeled by certain elements of the non-invertible algebra in a family-independent way. 
We extended this analysis to the case where each generation of fermions has been assigned to different elements. 
As an illustrative example, we deal with Ma model with the Ising fusion rule. Then, the neutrino mass matrix possesses the one-zero textures such as type-B and -C in Ref.~\cite{Barreiros:2018ndn}, providing peculiar predictions satisfying the neutrino oscillation data.

We focused on simple radiative neutrino mass models with an emphasis on the mechanism in section \ref{subsub301}, but it is fascinating to extend our analysis into other phenomenological models utilizing the other loop breaking scenarios presented in sections \ref{subsub302} - \ref{subsub303} and to elucidate phenomenological consequences of non-invertible selection rules. 
As discussed in Refs.~\cite{Kaidi:2024wio, Heckman:2024obe, Funakoshi:2024uvy}, the selection rule can be violated at the one-loop level. Indeed, in most of the models we have proposed, the terms breaking the selection rule are generated at one loop. 
However, in model 2, it appears only at the three-loop level, which is intriguing. 
This behavior arises due to the specific relations among the gauge charges and fermion numbers. This direction will be further explored in future work.

\acknowledgments

This work was supported by JSPS KAKENHI Grant Numbers JP23K03375 (T.K.),  Zhongyuan Talent (Talent Recruitment Series) Foreign Experts Project (H.Okada) and JP25H01539 (H.Otsuka).

\appendix

\section{Selection rules}
\label{app}

In this section, we summarize three non-invertible fusion algebras utilized in this paper:
\begin{enumerate}
    \item Fibonacci fusion algebra
\begin{align}
    \mathbb{I}\otimes \mathbb{I} = \mathbb{I},\quad
    \mathbb{I}\otimes \tau = \tau \otimes \mathbb{I} = \tau,\quad    
    \tau \otimes \tau = \mathbb{I} \oplus \tau.
\end{align}

\item Ising fusion algebra:
\begin{align}
    \mathbb{I}\otimes \mathbb{I} = \mathbb{I},\quad
    \mathbb{I}\otimes x = x \otimes \mathbb{I} = x,\quad    
    \epsilon \otimes \epsilon = \mathbb{I},\quad
    \epsilon \otimes \sigma = \sigma \otimes \epsilon = \sigma,\quad
    \sigma \otimes \sigma = \mathbb{I} \oplus \epsilon\,,
\end{align}
with $x=\{\mathbb{I},\epsilon,\sigma\}$.

\item  $\mathbb{Z}_3$ Tambara-Yamagami fusion algebra:
\begin{align}
    \eta^3 = \mathbb{I},\quad
    \eta \otimes {\cal N} = {\cal N},\quad
    {\cal N}\otimes {\cal N} = \mathbb{I} + \eta + \eta^2.
\end{align}

\end{enumerate}

The former two algebras are used to describe non-Abelian anyons in fractional quantum Hall states~\cite{Moore:1991ks,Nayak:1996irt,Read:1998ed,Slingerland:2001ea}. 
The latter one appears in the 3-states Potts model and the $\mathbb{Z}_3$ parafermion theory. 
In the following, we briefly review one of the origins of non-invertible fusion algebras from the viewpoint of field theory, following Refs.~\cite{Kobayashi:2024yqq,Funakoshi:2024uvy}.

Let us suppose that four-dimensional (4D) massless modes have the $\mathbb{Z}_M$ charges:
\begin{align}
    \varphi_k \rightarrow g^k \varphi_k,
\end{align}
with the $\mathbb{Z}_M$ generator $g=e^{2\pi i/M}$. For instance, compactifications of higher-dimensional Yang-Mills theory on torus with magnetic fluxes lead to the 4D effective field theory with $\mathbb{Z}_M$ symmetry \cite{Abe:2009vi,Berasaluce-Gonzalez:2012abm,Marchesano:2013ega}. 
When we extend this analysis to toroidal orbifold with $\mathbb{Z}_2$ twist, 4D $\mathbb{Z}_2$ invariant models are described as a linear combination of $\varphi_k$ and $\varphi_{M-k}$ \cite{Abe:2008fi}, i.e.,
\begin{align}
    \phi_k = \varphi_k + \varphi_{M-k},
\end{align}
up to a normalization factor. This $\mathbb{Z}_2$ invariant mode $\phi_k$ has no definite charge under the $\mathbb{Z}_M$ symmetry, but it can be labeled by a class under $\mathbb{Z}_2$ gauging of $\mathbb{Z}_M$ symmetry~\cite{Kobayashi:2024yqq,Funakoshi:2024uvy}, as introduced below.

Let us introduce the automorphism of $\mathbb{Z}_M$:
\begin{align}
    eg^k e^{-1}= g^k,\qquad
    rg^k r^{-1} = g^{-k}=g^{M-k},
\end{align}
which leads to the following classes:
\begin{align}
    [g^k]=\{hg^kh^{-1} ~|~ h=e, r \}.
\end{align}
Since such a class has both $k$ and $M-k$ charges, one can assign the field $\phi_k$ for the class $[g^k]$. 
Then, the product of two classes $[g^{k_1}]$ and $[g^{k_2}]$ obeys the following multiplication rule:
\begin{align}
\label{eq:multi-law}
    [g^{k_1}][g^{k_2}]=[g^{k_1+k_2}]+[g^{{k_1}-{k_2}}],
\end{align}
which is different from the multiplication law of the $\mathbb{Z}_M$ symmetry:
\begin{align}
    g^{k_1}g^{k_2}=g^{k_1+k_2}.
\end{align}
From the multiplication law of two classes \eqref{eq:multi-law}, two point couplings are allowed if 
\begin{align}
    \pm k_1 \pm k_2 = 0 ~~~({\rm mod~~}M)
\end{align}
is satisfied. Hence, the mass and kinetic terms are allowed for the same class $[k^1]=[k^2]$. 
In a similar way, one can extend to the $n$-point couplings $\phi_1\phi_2\cdots\phi_n$ each which belongs to the class $[g^{k_i}]$ $(i=1,2,...,n)$. The $n$-point couplings are allowed if
\begin{align}
    \sum_{i=1}^n \pm k_i  =0 ~~~({\rm mod~~}M)
\end{align}
is satisfied. 
Note that one can realize the Fibonacci fusion algebra for $M=3$ and the Ising fusion algebra for $M=4$. 
We restrict ourselves to three fusion algebras in this paper, but the other section rules with different $M$ can be applied, following the analysis of this paper. 
Furthermore, other selection rules are also derived in Ref.~\cite{Dong:2025pah}. 
It is interesting to apply these non-invertible selection rules to phenomenological models, which is beyond this paper. 

\bibliography{references}{}
\bibliographystyle{JHEP}

\end{document}